\documentclass[preprint,amsmath,amssymb,aps,pra,longbibliography]{revtex4-1}

\usepackage[margin=0.75in, top=1in, bottom=0.85in]{geometry}

\usepackage[final,letterspace=-0]{microtype}
\usepackage[colorlinks=true, linkcolor=blue, urlcolor=blue, citecolor=blue, anchorcolor=blue]{hyperref}

\usepackage{mathpazo} 
\usepackage{upgreek}

\usepackage{graphicx}
\usepackage{dcolumn}
\usepackage{bm}
\usepackage{helvet}
\usepackage{comment}
\usepackage{ulem}

\begin{document}

\title{Broadband entangled-photon omni-resonance in a planar optical cavity}

\author{Bryan L. Turo$^{1}$, Layton A. Hall$^{1,2}$, Bahaa E. A. Saleh$^{1}$, and Ayman F. Abouraddy$^{1,}$}
\email{Corresponding authors: raddy@creol.ucf.edu}
\affiliation{$^{1}$CREOL, The College of Optics \& Photonics,University of Central Florida, Orlando, Florida 32816, USA}
\affiliation{$^{2}$Materials Physics and Applications - Quantum Division, Los Alamos National Laboratory, Los Alamos, NM 87545, USA}

\begin{abstract}
Resonant field enhancement in an optical cavity is a promising pathway towards realizing optical nonlinearities at the few-photon level. This quest is hampered by inevitable narrowing of the resonant linewidth as the cavity finesse is increased, which necessitates striking a compromise between the magnitude of the field enhancement and the bandwidth over which it is harnessed. This difficulty is exacerbated for broadband entangled-photon pairs, which are typically frequency-anticorrelated, so that the two photons cannot be simultaneously admitted to a cavity except when the degenerate wavelength coincides with a cavity resonance. Here we show that introducing judicious angular dispersion into single-photon and entangled-photon states before incidence on a planar Fabry-P{\'e}rot (FP) cavity renders these non-classical fields omni-resonant: the entire spectrum is coupled to a single longitudinal cavity mode. Making use of a planar FP cavity of finesse $\approx100$, resonant linewidth $\approx0.3$~nm, and free spectral range $\approx22$~nm in the near-infrared, we couple single-photon states and frequency-anticorrelated entangled-photon states of 20-nm bandwidth to a broadband achromatic resonance associated with a single underlying narrowband longitudinal FP-cavity mode -- thereby preserving the entangled spectral structure. In general, pre-conditioning the optical field by introducing angular dispersion enables coupling non-classical states of light to a single longitudinal cavity mode, even if the field bandwidth far exceeds the resonant linewidth, or even exceeds the cavity free-spectral-range. These results pave the way to broadband resonant interactions with non-classical states of light in photon-starved applications.
\end{abstract}

\maketitle


Resonant field enhancement in optical cavities is crucial for a wide span of applications, from linear absorption (e.g., coherent perfect absorption \cite{Chong10PRL,Wan11S,Zhang12Light,Pye17OL,Baranov17NRM} and nonlinear optical effects \cite{Turner08OE,Hashemi09PRA,Lin16Optica,Ji17Optica,Koshelev20Science}, to ring-down spectroscopy \cite{Berden00IRPC,Ball03CR,Maity20AC} and gravitational wave detection \cite{Wise04CQG}. Inevitably, a dichotomy arises between the resonant field enhancement produced by light-recycling in a high-finesse optical cavity and the spectral linewidth over which it can be harnessed \cite{SalehBook07}. This fundamental trade-off between the cavity finesse and resonant linewidth limits the effectiveness of planar Fabry-P{\'e}rot (FP) cavities in broadband applications \cite{Ismail16OE}. This constraint poses particular challenges for broadband or multi-mode quantum optical applications, which nevertheless stand to benefit most from resonant field enhancement. The problem is particularly acute for entangled photons produced by spontaneous parametric downconversion (SPDC) \cite{Louisell61PR,Gordon63PR}, where energy conservation leads to frequency entanglement between the generated signal-idler photon pairs. Two key features underpin this challenge. First, the large bandwidth of SPDC photons -- which enables applications requiring precise photon synchronization \cite{Zukowski93PRL,Pan98PRL,Flamini18RPP,Lu22RMP,Azuma23RMP} or quantum spectroscopy \cite{Dorfman16PMP,Varnavski20JACS,Raymer21Optica,Landes21OE,Eshun21JACS,Eshun22ACR} -- vastly exceeds the resonant linewidth of a high-finesse cavity. Second, the frequencies of the two entangled photons are typically anticorrelated;  when one photon is red-shifted with respect to the degenerate frequency, the other is blue-shifted \cite{Kwiat95PRL,SalehBook07}. As such, if one photon couples to a cavity resonance, the other will be rejected; except at exact degeneracy when the two photons have the same wavelength. This is unfortunate because resonant field enhancement is likely indispensable for ultimately reaching the nonlinear regime at the single-photon or few-photon level \cite{Munro05NJP,Venkataraman13NP,Fryett15OE,Maser16NP,Staunstrup24NC,Gu25APR}. 

Given the central role played by optical cavities in many photonics applications, significant efforts have focused on broadening the resonant linewidth over which field buildup occurs -- without reducing the cavity finesse. One approach known as `white-light cavities' introduces an active medium into the cavity to provide anomalous dispersion, which can render the cavity round-trip phase independent of wavelength \cite{Wicht97OC,Rinkleff05PS}. To date, atomic \cite{Pati07PRL,Wu08PRA} and nonlinear \cite{Yum13JLT} resonances have been exploited for this purpose. However, such approaches have yielded $\sim$~MHz linewidth broadening, which is limited by the resonance linewidth of the active species. Exploiting linear passive components, whether prisms, diffraction grating \cite{Wise05PRL}, or chirped Bragg mirrors \cite{Yum13OC} cannot provide the requisite anomalous dispersion. Alternate approaches include enlarging the cavity size to increase the density of states \cite{Savchenkov06OL,Strekalov15OL}, nested cavities \cite{Kotlicki14OL}, among other possibilities \cite{Bliokh08RMP}.

We have recently pursued an alternative strategy that eschews modifications of the cavity itself, and instead relies on pre-conditioning the broadband incoming field \cite{Shabahang17SR} (see a related theoretical proposal in Ref.~\cite{Lin15OL}). This technique exploits the intrinsic angular dispersion (AD) \cite{Torres10AOP,Fulop10Review} underpinning conventional FP cavity resonances: the resonant wavelength associated with a particular longitudinal resonance at normal incidence blue-shifts at oblique incidence \cite{SalehBook07}. Therefore, by introducing AD into the incident field, whereby each wavelength is directed at a prescribed incident angle, the entire bandwidth of the angularly tailored field can be coupled to the cavity and resonate within it. We refer to this configuration as `omni-resonant' \cite{Shabahang17SR,Shabahang19OL,Shiri20OL,Shiri20APLP}: resonance is achieved over a continuous spectrum whose bandwidth not only vastly exceeds the resonant linewidth \cite{Villinger21AOM}, but may also exceed the cavity free spectral range (FSR) \cite{Hall25LPR}. Crucially, the bandwidth of this achromatic resonance is independent of the cavity finesse, so that the link is severed between the cavity photon lifetime and the resonant linewidth \cite{Shabahang17SR}.

Here we show that omni-resonance can be harnessed in quantum optics to efficiently couple broadband single-photon states and entangled-photon states to an achromatic resonance in a high-finesse planar FP cavity by introducing judicious AD into the quantum field. We experimentally verify this effect using entangled photon pairs produced via SPDC, making use of a planar FP cavity of finesse $\mathcal{F}\approx100$, a resonant linewidth of $\delta\lambda\approx0.3$~nm, and an FSR of $\sim22$~nm in the near-infrared. Without modifying the FP cavity or impacting its finesse, the broadband photons (of bandwidth $\approx20$~nm) couple to a single underlying narrow-linewidth longitudinal resonance of the FP cavity. Measuring the transmission through the FP cavity under both conventional resonant and omni-resonant conditions, single-photon states and frequency-anticorrelated entangled photon pairs of bandwidth $\sim20$~nm are coupled simultaneously to the same longitudinal resonance, and we confirm that the characteristic joint spectral structure of the entangled photons is preserved. Our results confirm that omni-resonance can be extended to quantum states of light, demonstrated here with single-photon and entangled-photon states. This raises the prospects for exploiting broadband resonant field enhancement in photon-starved applications in nonlinear spectroscopy, enhanced linear and nonlinear absorption of weak light fields, and quantum imaging and sensing.

\clearpage
\section*{Results}

\begin{figure*}[t!]
\centering
\includegraphics[width=18cm]{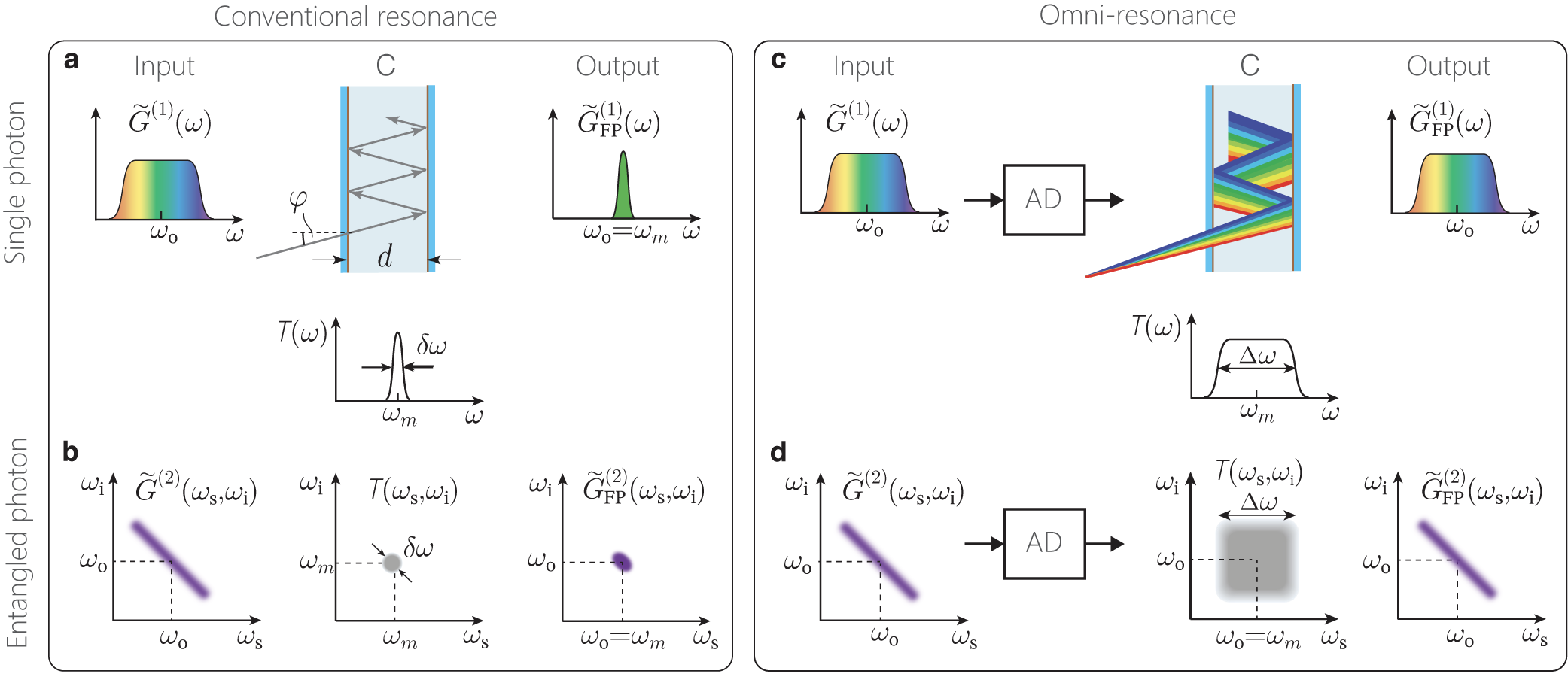}
\caption{\textbf{Conventional narrowband resonance versus broadband omni-resonance for single-photon and entangled-photon states.} (a) A broadband collimated single-photon state $\widetilde{G}^{(1)}(\omega)$ incident on a planar FP cavity is filtered by the resonant spectral transfer function $T(\omega)$ corresponding to a single conventional FP resonance of linewidth $\delta\omega$ at frequency $\omega_{m}$, $\widetilde{G}^{(1)}_{\mathrm{FP}}(\omega)$. (b) The biphoton spectrum $\widetilde{G}^{(2)}(\omega_{\mathrm{s}},\omega_{\mathrm{i}})$ for broadband frequency-anticorrelated entangled photons is also filtered by the FP cavity with two-photon resonant spectral transfer function $T_{2}(\omega_{\mathrm{s}},\omega_{\mathrm{i}})=T(\omega_{\mathrm{s}})T(\omega_{\mathrm{i}})$, to yield $\widetilde{G}_{\mathrm{FP}}^{(2)}(\omega_{\mathrm{s}},\omega_{\mathrm{i}})=T_{2}(\omega_{\mathrm{s}},\omega_{\mathrm{i}})\widetilde{G}^{(2)}(\omega_{\mathrm{s}},\omega_{\mathrm{i}})$. Both photons couple to the cavity only if the degenerate condition $\omega_{\mathrm{s}}=\omega_{\mathrm{i}}=\omega_{\mathrm{o}}$ coincides with a cavity resonance $\omega_{\mathrm{o}}=\omega_{m}$. Otherwise only one photon is coupled to the FP cavity while the other is rejected. (c,d) In the omni-resonant condition, pre-conditioning the incident light by introducing angular dispersion (AD) enables coupling the photon to a broadband achromatic resonance of the same FP cavity in (a,b). The omni-resonant transfer function now has a broad bandwidth $\Delta\omega\gg\delta\omega$, and may even exceed its FSR. (c) The single-photon and (d) the biphoton spectra are coupled in their entirety to the FP cavity.}
\label{Fig:SingleP_EntangledP_OmniRes}
\end{figure*}

\textbf{Coupling single-photon and entangled-photon states to a planar FP cavity.} Consider a broadband, collimated, single-photon state incident on a planar FP cavity (Fig.~\ref{Fig:SingleP_EntangledP_OmniRes}a). Most of the one-photon spectrum $\widetilde{G}^{(1)}(\omega)$ is reflected except for a narrow spectrum $\widetilde{G}^{(1)}_{\mathrm{FP}}(\omega)$ that couples to a longitudinal resonance of linewidth $\delta\omega$ at frequency $\omega_{m}$ (integer $m$ is the resonance order). When an entangled-photon (or biphoton, for brevity) state $\widetilde{G}^{(2)}(\omega_{\mathrm{s}},\omega_{\mathrm{i}})$ is incident on this cavity (Fig.~\ref{Fig:SingleP_EntangledP_OmniRes}b), similar spectral filtering occurs for \textit{each} photon to yield $\widetilde{G}^{(2)}_{\mathrm{FP}}(\omega_{\mathrm{s}},\omega_{\mathrm{i}})$; here $\omega_{\mathrm{s}}$ and $\omega_{\mathrm{i}}$ are the signal and idler frequencies, respectively. Consider a frequency-anticorrelated biphoton where $\omega_{\mathrm{s}}+\omega_{\mathrm{i}}=2\omega_{\mathrm{o}}$, where $\omega_{\mathrm{o}}$ is the degenerate frequency (half the pump frequency), which is typical for SPDC with a spectrally narrow pump \cite{SalehBook07}. In the degenerate configuration, we have  $\omega_{\mathrm{s}}=\omega_{\mathrm{i}}=\omega_{\mathrm{o}}$. Consequently, when the degenerate frequency coincides with a cavity resonance $\omega_{\mathrm{o}}=\omega_{m}$, both photons are admitted to the cavity at that frequency (Fig.~\ref{Fig:SingleP_EntangledP_OmniRes}b). However, if instead $\omega_{\mathrm{o}}\neq\omega_{m}$, then the two photons cannot be coupled to the cavity simultaneously. Rather, when the signal photon is coupled to the cavity at frequency $\omega_{\mathrm{s}}=\omega_{m}$, then the idler photon at frequency $\omega_{\mathrm{i}}=2\omega_{\mathrm{o}}-\omega_{m}\neq\omega_{m}$ is rejected; and vice versa. The challenge here is to facilitate coupling a broadband single-photon state (Fig.~\ref{Fig:SingleP_EntangledP_OmniRes}c) or a frequency-anticorrelated biphoton state (Fig.~\ref{Fig:SingleP_EntangledP_OmniRes}d) to a single longitudinal resonance without modifying the cavity itself, thereby maintaining its finesse. We aim to pre-condition the incident optical field to couple the photons (without spectral filtering) to a spectrally broad achromatic resonance of bandwidth $\Delta\omega$ that is nevertheless underpinned by a single conventional longitudinal resonance of linewidth $\delta\omega$, with $\Delta\omega\gg\delta\omega$ (indeed, $\Delta\omega$ may even exceed the cavity FSR). We refer to these scenarios as `single-photon omni-resonance' (Fig.~\ref{Fig:SingleP_EntangledP_OmniRes}c) and `entangled-photon omni-resonance' (Fig.~\ref{Fig:SingleP_EntangledP_OmniRes}d), respectively.

\begin{figure*}[t!]
\centering
\includegraphics[width=9cm]{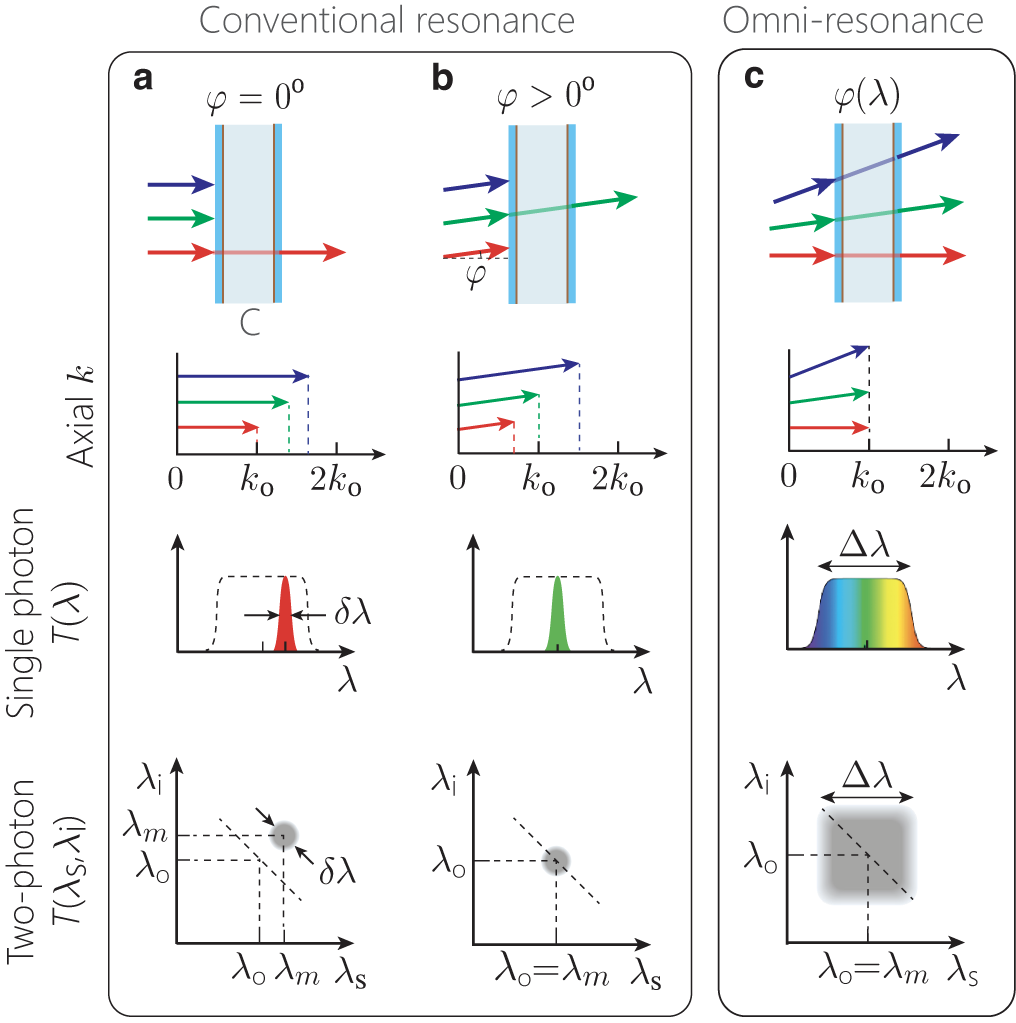}
\caption{\textbf{Transforming a conventional narrowband resonance into a broadband achromatic resonance in the omni-resonant configuration.} (a-c) Three broadband field configurations incident on a planar FP cavity: (a) collimated light at normal incidence; (b) obliquely incident collimated light at an angle $\varphi$ with the cavity normal, whereupon the resonant wavelength blue-shifts; and (c) angularly dispersed light with wavelength-dependent incident angle $\varphi(\lambda)$. The cases (a, b) correspond to the conventional resonant condition, whereas (c) corresponds to the broadband omni-resonant condition. The first row is a schematic of the incident-field configuration. The second row depicts the wave vectors in the cavity, the axial components of which must match certain values to resonate. The third row is the single-photon spectral transfer function $T(\lambda)$ of width $\delta\lambda$ overlaid with the input spectrum of width $\Delta\lambda$ (dashed spectrum). In (c) the bandwidth of the achromatic resonance is broadened to coincide with $\Delta\lambda$. The fourth row is the two-photon spectral transfer function $T_{2}(\lambda_{\mathrm{s}},\lambda_{\mathrm{i}})=T(\lambda_{\mathrm{s}})T(\lambda_{\mathrm{i}})$, which is the portion of the biphoton spectrum that resonates with the cavity. The resonant wavelength $\lambda_{m}$ differs from the degenerate wavelength $\lambda_{\mathrm{o}}$ in (a), but the two frequencies coincide in (b). Overlaid with the two-photon spectral transmission is a typical biphoton frequency-anticorrelated spectrum for reference (the faint anti-diagonal linear feature).}
\label{Fig:OmniResConcept}
\end{figure*}

\textbf{Concept of omni-resonance.} How can pre-conditioning a broadband optical field enable coupling to an achromatic resonance? Consider a planar FP cavity of length $d$ and refractive index $n$. A free-space wavelength $\lambda$ resonates with this cavity when the axial component of the associated wave vector in free space is $k_{m}=m\tfrac{\pi}{nd}$, which corresponds at normal incidence to $\lambda=\lambda_{m}=\tfrac{2nd}{m}$, where integer $m$ is the resonance order (Fig.~\ref{Fig:OmniResConcept}a). Oblique incidence from free space at an external angle $\varphi$ with the cavity normal blue-shifts the resonant wavelength to maintain this axial-wave-number constraint (Fig.~\ref{Fig:OmniResConcept}b): the new resonant wavelength at oblique incidence is $\lambda_{m}(\varphi)=\tfrac{2nd}{m}\sqrt{1-\tfrac{1}{n^{2}}\sin^{2}\varphi}$. Consequently, if each wavelength in the spectrum is directed to the cavity at a judiciously selected incident angle $\varphi(\lambda)$, then the entire spectrum resonates simultaneously (Fig.~\ref{Fig:OmniResConcept}c), which we refer to as omni-resonance. The spectral transfer function $T(\lambda)$ for classical light and single-photon states in the omni-resonant configuration is thus spectrally flat over a large bandwidth.

Not only does omni-resonance apply to classical and single-photon fields, but it also extends to biphoton fields. Consider a collimated biphoton state $|\Psi\rangle=\iint\!d\omega_{\mathrm{s}}d\omega_{\mathrm{i}}\widetilde{\psi}(\omega_{\mathrm{s}},\omega_{\mathrm{i}})|1_{\omega_{\mathrm{s}}},1_{\omega_{\mathrm{i}}}\rangle$ incident on an FP cavity having a single-photon spectral transfer function $T(\lambda)$; here $|1_{\omega}\rangle$ is the one-photon Fock state at frequency $\omega$, $\langle1_{\omega}|1_{\omega'}\rangle=\delta(\omega-\omega')$, and we make use of the normalization $\iint\!d\omega_{\mathrm{s}}d\omega_{\mathrm{i}}|\widetilde{\psi}(\omega_{\mathrm{s}},\omega_{\mathrm{i}})|^{2}=1$. The biphoton spectrum before the FP cavity is $\widetilde{G}^{(2)}(\omega_{\mathrm{s}},\omega_{\mathrm{i}})=|\widetilde{\psi}(\omega_{\mathrm{s}},\omega_{\mathrm{i}})|^{2}$, which becomes after the cavity $\widetilde{G}_{\mathrm{FP}}^{(2)}(\omega_{\mathrm{s}},\omega_{\mathrm{i}})=T_{2}(\omega_{\mathrm{s}},\omega_{\mathrm{i}})\widetilde{G}^{(2)}(\omega_{\mathrm{s}},\omega_{\mathrm{i}})$; here $T_{2}(\omega_{\mathrm{s}},\omega_{\mathrm{i}})=T(\omega_{\mathrm{s}})T(\omega_{\mathrm{i}})$ is the biphoton spectral transfer function of the cavity.

Consider a frequency-anticorrelated biphoton state, $\widetilde{G}^{(2)}(\omega_{\mathrm{s}},\omega_{\mathrm{i}})\propto\widetilde{g}(\omega_{\mathrm{s}}+\omega_{\mathrm{i}}-2\omega_{\mathrm{o}})$, where $\widetilde{g}(\omega)$ is a narrow spectral function and $\omega_{\mathrm{o}}$ is the degenerate frequency, then the two photons cannot resonate simultaneously except when two conditions are satisfied: (1) the photons are in a degenerate configuration $\omega_{\mathrm{s}}=\omega_{\mathrm{i}}=\omega_{\mathrm{o}}$; and (2) the degenerate frequency coincides with a cavity resonance, $\omega_{\mathrm{o}}=\omega_{m}$. If $\omega_{\mathrm{o}}\neq\omega_{m}$ at normal incidence, only one photon can resonate at a time, while the other photon is rejected (Fig.~\ref{Fig:OmniResConcept}a). However, the incident angle $\varphi$ can be varied to achieve $\omega_{m}(\varphi)=\omega_{\mathrm{o}}$, whereupon the narrow bandwidth $\delta\omega$ associated with the $m^{\mathrm{th}}$ longitudinal mode coincides with the degenerate frequency, which allows for the two photons to resonate simultaneously over the narrow resonant linewidth (Fig.~\ref{Fig:OmniResConcept}b). Now, if each wavelength in the biphoton state is directed to the cavity at an appropriate angle $\varphi(\lambda)$, as in the single-photon state described above (Fig.~\ref{Fig:OmniResConcept}c), then the entire bandwidth of the biphoton state resonates, and the frequency-anticorrelated biphoton spectrum is preserved.

\begin{figure*}[t!]
\centering
\includegraphics[width=17.6cm]{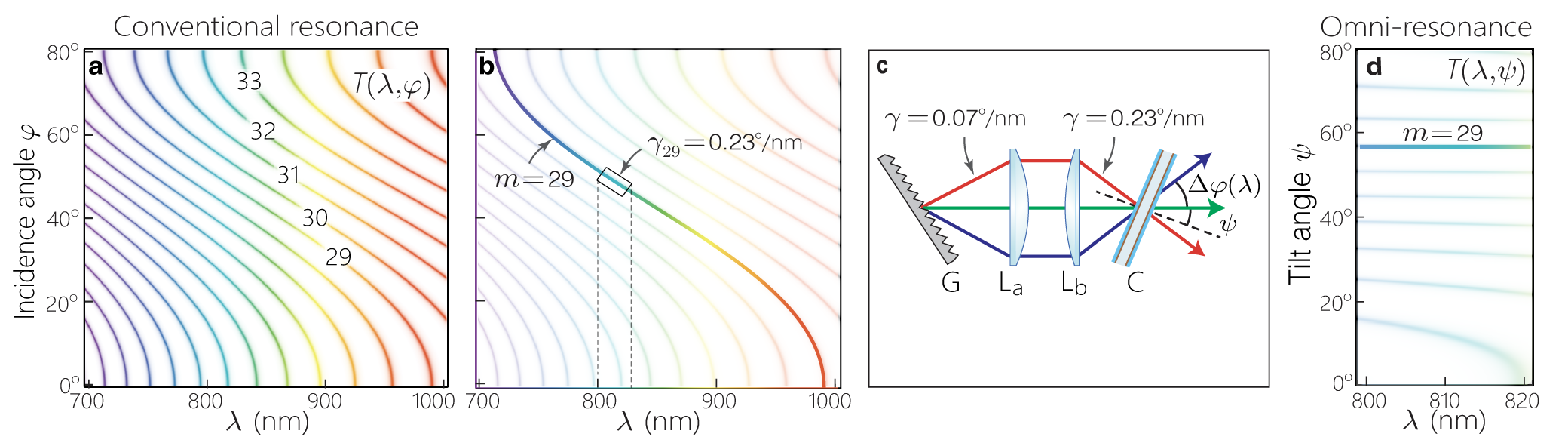}
\caption{\textbf{Realizing an achromatic resonance.} (a) The angular dependence of the spectral transfer function $T(\lambda,\varphi)$ of the planar FP cavity used in our experiments, where $\varphi$ is the external incident angle with the cavity normal of a collimated field; see Fig.~\ref{Fig:OmniResConcept}a,b. We plot $T(\lambda,\varphi)$ with a reduced finesse of $\mathcal{F}=10$ for clarity. (b) A single conventional resonance from (a) corresponding to $m=29$ is isolated, and we highlight the spectral range utilized for omni-resonance. (c) The optical system to introduce AD into the incident field. G: Diffraction grating; L$_{\mathrm{a}}$, L$_{\mathrm{b}}$: lenses of focal lengths 200~mm and 55~mm, respectively; C: planar FP cavity. (d) The spectral transfer function $T(\lambda,\psi)$ of the achromatic resonance after introducing AD into the conventional resonance in (b). Here $\psi$ (the cavity tilt angle) is the angle made by $\lambda=810$~nm with the cavity normal, corresponding to the dashed line in (c). The direction of this wavelength defines the optical axis. Other wavelengths travel at an angle $\Delta\varphi(\lambda)$ with the respect to this optical axis, thereby making an angle $\psi+\Delta\varphi(\lambda)$ with the cavity normal.}
\label{Fig:Cavity}
\end{figure*}

\textbf{Realizing omni-resonance.} What pre-conditioning of the optical field is required to achieve omni-resonance? The spectral transfer function $T(\lambda,\varphi)$ of a symmetric FP cavity for a collimated, obliquely incident broadband field is $T(\lambda,\varphi)=1\big/\{1 +\bigl(\frac{2\mathcal{F}}{\pi}\bigr)^2 \sin^2\!\frac{\chi}{2}\}$, where $\mathcal{F}=\frac{\pi \sqrt{R}}{1-R}$ is the cavity finesse, $R$ is the mirror reflectivity, the roundtrip phase is $\chi(\lambda,\varphi)=\frac{4\pi d}{\lambda}\,\sqrt{n^2 - \sin^{2}\varphi}$ (we ignore the reflection phases from the two cavity mirrors), and $\varphi$ is the external incident angle from free space with respect to the cavity normal. Resonances occur whenever $\chi(\lambda,\varphi)=2\pi m,$ where $m$ is an integer (the resonance order), a condition that yields a discrete set of resonant wavelengths $\lambda_{m}(\varphi)=\tfrac{2d}{m}\sqrt{n^{2}-\sin^{2}\varphi}$ for any incident angle, with the resonant wavelength blue-shifting with increasing $\varphi$. We plot $T(\lambda,\varphi)$ in Fig.~\ref{Fig:Cavity}a for the cavity used in our experiments, with $R=0.97$, $d\approx9.88$~$\mu$m, $n=1.45$, and thus $\mathcal{F}\approx 100$.

Consider a single FP resonance (isolated from all the others in Fig.~\ref{Fig:Cavity}a) as illustrated in Fig.~\ref{Fig:Cavity}b for $m=29$. By varying $\varphi$, the resonant wavelength can be tuned from $\lambda_{m}(0^{\circ})=\tfrac{2nd}{m}$ at normal incidence to $\lambda_{m}(90^{\circ})=\tfrac{2d}{m}\sqrt{n^{2}-1}$ at glancing incidence; here $\lambda_{29}(0^{\circ})=988$~nm and $\lambda_{29}(90^{\circ})=710$~nm. If AD is introduced into the incident field, so that the wavelength $\lambda$ is directed to the cavity at an angle $\varphi_{m}(\lambda)=\sin^{-1}\left\{n\sqrt{1-(\tfrac{\lambda}{\lambda_{m}})^{2}}\right\}$, then the entire spectrum extending from $\lambda_{m}(0)$ to $\lambda_{m}(90^{\circ})$ can resonate with the $m^{\mathrm{th}}$ longitudinal resonance. In principle, the omni-resonant bandwidth is $\Delta\lambda_{m}=\lambda_{m}(0^{\circ})-\lambda_{m}(90^{\circ})$, which can be quite extensive; here $\Delta\lambda_{29}=278$~nm. The cavity parameters remain unchanged, which preserves its finesse. Nevertheless, the requisite AD is a highly nonlinear function of $\lambda$, the bandwidth $\Delta\lambda_{m}$ is very large, and the numerical aperture necessary reaches unity, all of which are extremely challenging to realize experimentally. We exploit here only a portion of the bandwidth ($\Delta\lambda\approx20$~nm) from the available resonant spectrum ($\Delta\lambda_{29}=278$~nm), as highlighted in Fig.~\ref{Fig:Cavity}b. 

Consider a portion of the spectrum centered at $\lambda_{\mathrm{o}}$ at which the $m^{\mathrm{th}}$ resonance is achieved at an incident angle $\varphi_{m}(\lambda_{\mathrm{o}})=\psi$. Neighboring wavelengths resonate when incident at an angle $\varphi_{m}(\lambda)=\psi+\Delta\varphi_{m}(\lambda)$, with $\Delta\varphi_{m}(\lambda_{\mathrm{o}})=0$. We expand $\Delta\varphi_{m}(\lambda)\approx\gamma_{m}(\lambda-\lambda_{\mathrm{o}})$ to first order in $\lambda-\lambda_{\mathrm{o}}$, where $\gamma_{m}=\tfrac{d\varphi_{m}}{d\lambda}\big|_{\lambda_{\mathrm{o}}}$ is the first-order (linear) AD coefficient for the $m^{\mathrm{th}}$ resonance at $\lambda_{\mathrm{o}}$. In our case with $m=29$ and $\lambda_{\mathrm{o}}=810$~nm, we have $\gamma_{29}\approx0.23^{\circ}$/nm, which is a relatively large value. This can be appreciated by comparing $\gamma_{29}$ to the AD produced in the first diffraction order at $\lambda_{\mathrm{o}}=810$~nm by a grating of groove density 1200~lines/mm, for incident and diffracted angles with the cavity normal of $\alpha\approx40.3^{\circ}$ and $19.2^{\circ}$, respectively, with the latter representing the subsequent optical axis (Fig.~\ref{Fig:Cavity}c). The AD produced with these parameters is $\gamma\approx0.07^{\circ}$/nm, which falls significantly short of the required value of $\gamma_{29}\approx0.23^{\circ}$/nm. Several approaches may be adopted to enhance the AD from the grating \cite{Villinger21AOM}. For simplicity, we make use here of an imaging configuration formed of two lenses L$_{\mathrm{a}}$ and L$_{\mathrm{b}}$ of focal lengths 200 and 55~mm, respectively, in an imaging configuration (Fig.~\ref{Fig:Cavity}c). The de-magnification in space is accompanied by a broadening of the angular spectrum, and thus enlargement of the AD coefficient by $\times3.29$. The field incident on the cavity now has an AD coefficient of $\gamma=\gamma_{29}=0.23^{\circ}$/nm as required for omni-resonance, with the wavelength $\lambda_{\mathrm{o}}=810$~nm incident at an angle $\psi=59^{\circ}$ with the cavity normal (we refer to $\psi$ as the cavity tilt angle). The conventional narrowband FP resonance of order $m=29$ in Fig.~\ref{Fig:Cavity}b is thus converted to the broadband achromatic resonance shown in Fig.~\ref{Fig:Cavity}d. At this value of the cavity tilt angle $\psi$ and AD coefficient $\gamma$, the FP resonances of other orders ($m\neq29$) are not necessarily rendered achromatic. Nevertheless, each FP resonance can be transformed in turn into a broadband achromatic resonance by selecting appropriate values of $\psi$ and $\gamma$ at the target wavelength. Further increasing the omni-resonant bandwidth requires adding higher-order terms to the expansion of $\Delta\varphi_{m}(\lambda)$ beyond the first-order term incorporated here, which requires a different approach for incorporating AD \cite{Shiri20OL,Hall25LPR}.

\textbf{Biphoton source.} The source of biphotons in our experiments is a 3-mm-thick BBO nonlinear crystal (NLC) pumped by a monochromatic laser at a wavelength $\lambda_{\mathrm{p}}\approx405$~nm in a degenerate, collinear, type-I SPDC configuration (Fig.~\ref{Fig:heralded_setup}a). The general biphoton state produced is \cite{Saleh00PRA}:
\begin{equation}
|\Psi\rangle=\iiiint\!dq_{\mathrm{s}}d\omega_{\mathrm{s}}dq_{\mathrm{i}}d\omega_{\mathrm{i}}\widetilde{\psi}(q_{\mathrm{s}},\omega_{\mathrm{s}};q_{\mathrm{i}},\omega_{\mathrm{i}})|1_{q_{\mathrm{s}},\omega_{\mathrm{s}}};1_{q_{\mathrm{i}},\omega_{\mathrm{i}}}\rangle,
\end{equation}
where $q_{\mathrm{s}}$ and $\omega_{\mathrm{s}}$ are the transverse wave number and frequency of the signal photon, respectively, and $q_{\mathrm{i}}$ and $\omega_{\mathrm{i}}$ are the corresponding quantities for the idler photon, $|1_{q,\omega}\rangle$ is the one-photon Fock state associated with the optical mode identified by $q$ and $\omega$, and the biphoton spectral probability amplitude is normalized such that $\iiiint\!dq_{\mathrm{s}}d\omega_{\mathrm{s}}dq_{\mathrm{i}}d\omega_{\mathrm{i}}|\widetilde{\psi}(q_{\mathrm{s}},\omega_{\mathrm{s}};q_{\mathrm{i}},\omega_{\mathrm{i}})|^{2}=1$. For a monochromatic pump (at a frequency $\omega_{\mathrm{p}}$) of transverse spatial profile $E_{\mathrm{p}}(x)$, we have:
\begin{equation}
\widetilde{\psi}(q_{\mathrm{s}},\omega_{\mathrm{s}};q_{\mathrm{i}},\omega_{\mathrm{i}})\propto\widetilde{E}_{\mathrm{p}}(q_{\mathrm{s}}+q_{\mathrm{i}})\delta(\omega_{\mathrm{p}}-\omega_{\mathrm{s}}-\omega_{\mathrm{i}})\widetilde{\xi}(q_{\mathrm{s}},\omega_{\mathrm{s}};q_{\mathrm{i}},\omega_{\mathrm{i}}),
\end{equation}
where $\widetilde{E}_{\mathrm{p}}(q)$ is the Fourier transform of $E_{\mathrm{p}}(x)$, and $\widetilde{\xi}$ is the longitudinal phase-matching function given by: $\widetilde{\xi}(q_{\mathrm{s}},\omega_{\mathrm{s}};q_{\mathrm{i}},\omega_{\mathrm{i}})\propto\mathrm{sinc}\left(\tfrac{L}{2\pi}\Delta\zeta\right)$, $L$ is the NLC length, $\zeta$ is the axial wave number, $\Delta\zeta=\zeta_{\mathrm{p}}-\zeta_{\mathrm{s}}-\zeta_{\mathrm{i}}$, $q_{j}^{2}+\zeta_{j}^{2}=k_{j}^{2}$, and $j=\mathrm{p},\mathrm{s},\mathrm{i}$. We plot in Fig.~\ref{Fig:heralded_setup}b the signal-photon spectral probability $\widetilde{G}^{(1)}(q_{\mathrm{s}},\lambda_{\mathrm{s}})=|\widetilde{\xi}(q_{\mathrm{s}},\omega_{\mathrm{s}};-q_{\mathrm{s}},2\omega_{\mathrm{o}}-\omega_{\mathrm{s}})|^{2}$. We restrict the bandwidth of our experiment to $\Delta\lambda\approx20$~nm, which then restricts the spatial bandwidth to $\Delta q_{\mathrm{s}}\approx 0.4$~rad/$\mu$m, and further spatial magnification of the signal photon (see below) reduces the spatial bandwidth to $\Delta q_{\mathrm{s}}\approx 0.04$~rad/$\mu$m, corresponding to an angular bandwidth of $0.3^{\circ}$. Larger magnification can further reduce $\Delta q_{\mathrm{s}}$ and thus approach an ideally collimated field. We hence drop $q_{\mathrm{s}}$ and $q_{\mathrm{i}}$ from the biphoton spectral probability amplitude, which is approximated as $\widetilde{\psi}(\omega_{\mathrm{s}},\omega_{\mathrm{i}})\propto\delta(2\omega_{\mathrm{o}}-\omega_{\mathrm{s}}-\omega_{\mathrm{i}})\widetilde{\xi}(0,\omega_{\mathrm{s}};0,\omega_{\mathrm{i}})$, where $\widetilde{\xi}(0,\omega_{\mathrm{s}};0,\omega_{\mathrm{i}})\propto\mathrm{sinc}(\tfrac{L}{2\pi}k_{2}\Omega^{2})$, and the biphoton spectrum is $\widetilde{G}^{(2)}(\omega_{\mathrm{s}},\omega_{\mathrm{i}})=|\widetilde{\psi}(\omega_{\mathrm{s}},\omega_{\mathrm{i}})|^{2}$ (Fig.~\ref{Fig:heralded_setup}c); here $\omega_{\mathrm{s}}=\omega_{\mathrm{o}}+\Omega$, $\omega_{\mathrm{i}}=\omega_{\mathrm{o}}-\Omega$, and $k_{2}$ is the group-velocity dispersion coefficient in the nonlinear crystal at $\omega_{\mathrm{o}}$.

\textbf{Setup for biphoton generation.} We produce single-photon states by detecting one photon (the idler) to herald the arrival of the other (the signal); see Fig.~\ref{Fig:heralded_setup}d. The setup for producing SPDC biphotons comprises a collimated, monochromatic pump laser (Coherent, Cube) at a wavelength 405~nm and $1/e^2$ width of $\approx1$~mm, which is weakly focused with a lens of focal length $f=100$~mm to a 3-mm-thick BBO crystal (Fig.~\ref{Fig:heralded_setup}e), which helps increase the heralded photon count. Because the Rayleigh range $z_{\mathrm{R}}=10.3$~mm for the focused pump is approximately $3.4\times$ the NLC length, we can assume that its beam width is approximately constant along the NLC. The biphotons are produced in a collinear, degenerate, type-I configuration with degenerate wavelength $\lambda_{\mathrm{o}}=2\lambda_{\mathrm{p}}=810$~nm. The pump is eliminated after the NLC via a bandpass filter of bandwidth 40~nm (from 780 to 820~nm), but the \textit{joint} spectrum is limited to $\Delta\lambda=20$~nm centered at the degenerate wavelength of $\lambda_{\mathrm{o}}=810$~nm. The output facet of the NLC is imaged via a pair of lenses L$_{1}$ and L$_{2}$ (focal lengths 50 and 500~mm, respectively) that provide $10\times$ magnification. The image plane is at the entrance to the omni-resonant pre-conditioning system, which is depicted in Fig.~\ref{Fig:Cavity}c, followed by the FP cavity.

\noindent{\textbf{Single-photon omni-resonance.}} To verify single-photon omni-resonance, we first spatially separate the signal and idler photons with two paths via a beam splitter placed immediately after the spectral filter (Fig.~\ref{Fig:heralded_setup}e). The idler photon is focused using a lens ($f=50$~mm) and collected via a single-mode fiber that delivers the idler to a single-photon detector D$_{1}$ to herald the signal photon. The signal photon is imaged to the pre-conditioning system before traversing the FP cavity, after which a lens ($f=50$~mm) collimates the spatially resolved spectrum emerging from the cavity. A multimode fiber is scanned along the resolved spectrum, corresponding to $\lambda_{\mathrm{s}}$, and the fiber delivers the signal photon to another single-photon detector D$_{2}$. We refer to the measurements of the signal-photon counts as as `singles', and to the corresponding measurements jointly with the idler as `heralded'. For reference, we plot in Fig.~\ref{Fig:heralded_setup}f the signal-photon spectrum $\widetilde{G}^{(1)}(\lambda_{\mathrm{s}})$ in absence of the FP cavity. Here the pre-conditioning system acts simply as a spectral analyzer. The wavelengths obtained by scanning the multimode fiber are calibrated by aligning a classical optical beam from a superluminescent diode with the signal photon and connecting the multimode fiber to an optical spectrum analyzer. 

We present the single-photon measurements in the resonant and omni-resonant configurations in Fig.~\ref{Fig:SinglePhotonData}. We first plot in Fig.~\ref{Fig:SinglePhotonData}a,b measurements in the conventional resonant configuration: in Fig.~\ref{Fig:SinglePhotonData}a the singles measurements $\widetilde{G}^{(1)}_{\mathrm{FP}}(\lambda_{\mathrm{s}},\varphi)$ and in Fig.~\ref{Fig:SinglePhotonData}b the heralded measurements $\widetilde{G}^{(1)}_{\mathrm{FP}}(\lambda_{\mathrm{s}},\varphi)$ for the signal photon spectrum after traversing the FP cavity. Here the cavity is placed before the pre-conditioning system, which then acts as a spectral analyzer. The spectral measurements are collected over a bandwidth $\Delta\lambda\approx20$~nm spanning the wavelengths from 800 to 820~nm, and the incident angle is tuned from $\varphi=0^{\circ}$ (normal incidence) to $\varphi=59^{\circ}$ in steps of $0.55^{\circ}$ after mounting the cavity on a motorized rotation stage. The spectral resolution of the spectral measurements is estimated to be $\delta\lambda\approx0.9$~nm (FWHM), which is limited by the diameter of the MMF collecting the signal photons. Comparing Fig.~\ref{Fig:SinglePhotonData}a to Fig.~\ref{Fig:SinglePhotonData}b reveals no basic differences between the singles and heralded measurements as expected. We plot beneath Fig.~\ref{Fig:SinglePhotonData}a,b sections through $\widetilde{G}^{(1)}_{\mathrm{FP}}(\lambda_{\mathrm{s}},\varphi)$ and $\widetilde{G}^{(2)}_{\mathrm{FP}}(\lambda_{\mathrm{s}},\varphi)$  corresponding to fixed incident angles $\varphi=0^{\circ}$, $10^{\circ}$, and $58^{\circ}$. Two attributes emerge clearly: the spectrum features discrete resonances; and the resonant wavelengths (associated with any resonant order) blue-shift with increasing $\varphi$. We also measure the resonant transmission spectra at $\varphi=0^{\circ}$, $10^{\circ}$, and $58^{\circ}$ using a classical source and acquired with an optical spectrum analyzer, which are plotted in the bottom panels in Fig.~\ref{Fig:SinglePhotonData}a as dashed curves, yielding a measured resonant linewidth of $\delta\lambda\approx0.3$~nm. Note that the measured single-photon spectra $\widetilde{G}^{(1)}_{\mathrm{FP}}(\lambda_{\mathrm{s}},\varphi)$ is broader than the classical spectra because of the finite spectral bandwidth associated with the scanning multimode fiber.

The single-photon measurements undergo a major change after switching to the omni-resonant configuration as a consequence of introducing AD into the signal-photon field. Here the cavity is placed after the pre-conditioning system as illustrated in Fig.~\ref{Fig:heralded_setup}e. The spectral measurements are again collected over the same bandwidth of $\Delta\lambda=20$~nm extending from 800 to 820~nm. In presence of pre-conditioning, we no longer have a collimated field with a well-defined incident angle on the cavity. Rather, the signal-photon field is angularly dispersed. We utilize as an identifier the cavity tilt angle $\psi$, which is the incident angle of the degenerate wavelength $\lambda_{\mathrm{o}}=810$~nm with the cavity normal. The value of AD introduced, $\gamma\approx0.23^{\circ}$/nm, corresponds to the intrinsic FP resonance AD for $m=29$ at $\psi\approx 59^{\circ}$. Once again, the cavity is mounted on a motorized rotation stage, and $\psi$ is varied in steps of $0.55^{\circ}$ from $0^{\circ}$ to $60^{\circ}$. We note a substantial change in the measured single-photon spectrum $\widetilde{G}_{\mathrm{FP}}^{(1)}(\lambda_{\mathrm{s}},\psi)$ in Fig.~\ref{Fig:SinglePhotonData}c from $\widetilde{G}_{\mathrm{FP}}^{(1)}(\lambda_{\mathrm{s}},\varphi)$ in Fig.~\ref{Fig:SinglePhotonData}a. Specifically, the slopes of the resonances have changed and are now flattened horizontally over an extended bandwidth for select values of $\psi$. This is brought out even clearer by examining section through $\widetilde{G}_{\mathrm{FP}}^{(1)}(\lambda_{\mathrm{s}},\psi)$ for fixed values of $\psi=0^{\circ}$, $10^{\circ}$, and $59^{\circ}$, which show a very different picture from the conventional resonant scenario. Here in the omni-resonant condition at the intended tilt angle $\psi=59^{\circ}$ we observe a broad, spectrally flat spectral transfer function over the full bandwidth of $\Delta\lambda=20$~nm. At other cavity tilt angles, the spectral transfer function extends over a more limited bandwidth, and even discrete resonances emerge at small values of $\psi$. Once again, the heralded signal-photon spectrum $\widetilde{G}^{(2)}(\lambda_{\mathrm{s}},\psi)$ in Fig.~\ref{Fig:SinglePhotonData}d resembles the corresponding singles measurements in Fig.~\ref{Fig:SinglePhotonData}c.

\textbf{Biphoton omni-resonance.} Biphoton omni-resonance is realized using a setup similar to that in Fig.~\ref{Fig:heralded_setup}, which we illustrate in Fig.~\ref{Fig:EntangledPhotonSetup}. The central difference is that the signal and idler photons are \textit{not} separated into different spatial paths prior to the AD-synthesis stage. Rather, both signal and idler are directed to AD-synthesis and together impinge on the FP cavity (Fig.~\ref{Fig:EntangledPhotonSetup}a). Only after the two photons emerge from the FP cavity are they separated by a beam splitter, as shown Fig.~\ref{Fig:EntangledPhotonSetup}b. Each photon is directed to a scanning multimode fiber connected to a single-photon detector along the spatially resolved spectrum. By scanning two fibers, one in the signal path and the other in the idler path, we capture the biphoton spectrum $\widetilde{G}_{\mathrm{FP}}^{(2)}(\lambda_{\mathrm{s}},\lambda_{\mathrm{i}})=\widetilde{G}^{(2)}(\lambda_{\mathrm{s}},\lambda_{\mathrm{i}})T_{2}(\lambda_{\mathrm{s}},\lambda_{\mathrm{i}})$, where $T_{2}(\lambda_{\mathrm{s}},\lambda_{\mathrm{i}})$ is the biphoton spectral transfer function of the FP cavity.

We start by measuring the joint spectrum in the conventional resonant configuration by placing the cavity before the pre-conditioning system. The biphoton spectrum after traversing the cavity is $\widetilde{G}_{\mathrm{FP}}^{(2)}(\lambda_{\mathrm{s}},\lambda_{\mathrm{i}};\varphi)=T_{2}(\lambda_{\mathrm{s}},\lambda_{\mathrm{i}};\varphi)\widetilde{G}^{(2)}(\lambda_{\mathrm{s}},\lambda_{\mathrm{i}})$
, where $T_{2}(\lambda_{\mathrm{s}},\lambda_{\mathrm{i}};\varphi)=T(\lambda_{\mathrm{s}},\varphi)T(\lambda_{\mathrm{i}},\varphi)$, the single-photon transfer function $T(\lambda_{\mathrm{s}},\varphi)$ is plotted in Fig.~\ref{Fig:EntPhotOmniResData}a (taken from Fig.~\ref{Fig:SinglePhotonData}a) and the biphoton spectrum before the cavity $\widetilde{G}^{(2)}(\lambda_{\mathrm{s}},\lambda_{\mathrm{i}})$ is plotted in Fig.~\ref{Fig:EntPhotOmniResData}b. In the first row of Fig.~\ref{Fig:EntPhotOmniResData}c-e we plot $T_{2}(\lambda_{\mathrm{s}},\lambda_{\mathrm{i}};\varphi)$ obtained from the product of the measured spectral transfer functions $T(\lambda_{\mathrm{s}},\varphi)$ and $T(\lambda_{\mathrm{i}},\varphi)$ from Fig.~\ref{Fig:EntPhotOmniResData}a at $\varphi=0$ (Fig.~\ref{Fig:EntPhotOmniResData}c), $10^{\circ}$ (Fig.~\ref{Fig:EntPhotOmniResData}d), and $58^{\circ}$ (Fig.~\ref{Fig:EntPhotOmniResData}e). The measured bandwidth is $\Delta\lambda=20$~nm, extending from 800 to 820~nm, which is centered at the degenerate wavelength of $\lambda_{\mathrm{o}}=810$~nm. We observe a single conventional FP resonance: $\lambda_{m}\approx817$~nm at $\varphi=0^{\circ}$ corresponding to $m=27$ (Fig.~\ref{Fig:EntPhotOmniResData}c); $\lambda_{m}\approx810$~nm at $\varphi=10^{\circ}$ corresponding to $m=27$ (Fig.~\ref{Fig:EntPhotOmniResData}d); and $\lambda_{m}\approx810$~nm at $\varphi=58^{\circ}$ corresponding to $m=29$ (Fig.~\ref{Fig:EntPhotOmniResData}e).

Because the spectral transfer function in Fig.~\ref{Fig:EntPhotOmniResData}c at $\varphi=0^{\circ}$ does \textit{not} overlap with the incident biphoton spectrum $\widetilde{G}^{(2)}(\lambda_{\mathrm{s}},\lambda_{\mathrm{i}})$, the signal and idler photons cannot be admitted simultaneously to the FP cavity. Rather, only one photon from the pair can resonate at a time. Therefore, the measured biphoton spectrum $\widetilde{G}_{\mathrm{FP}}^{(2)}(\lambda_{\mathrm{s}},\lambda_{\mathrm{i}};0^{\circ})$ after the FP cavity is essentially zero for all wavelength pairs. At $\varphi=10^{\circ}$ and $\varphi=58^{\circ}$, the spectral transfer function has its resonant peak at $\lambda_{\mathrm{s}}=\lambda_{\mathrm{i}}=\lambda_{\mathrm{o}}$, so that the signal and idler photons can both be admitted to the cavity in the vicinity of $\lambda_{\mathrm{o}}$. The measured biphoton spectrum $\widetilde{G}_{\mathrm{FP}}^{(2)}(\lambda_{\mathrm{s}},\lambda_{\mathrm{i}};\varphi)$ after the FP cavity corresponds to the overlap between $T(\lambda_{\mathrm{s}},\lambda_{\mathrm{i}};\varphi)$ and $\widetilde{G}^{(2)}(\lambda_{\mathrm{s}},\lambda_{\mathrm{i}})$, and only a portion of the biphoton state centered at $\lambda_{\mathrm{o}}$ is admitted to the cavity.

We now move the cavity after the pre-conditioning system as shown in Fig.~\ref{Fig:EntangledPhotonSetup}b. Both photons undergo pre-conditioning before traversing the cavity, so that both photons are independently in an omni-resonant configuration. The biphoton spectrum after the cavity is $\widetilde{G}_{\mathrm{FP}}^{(2)}(\lambda_{\mathrm{s}},\lambda_{\mathrm{i}};\psi)=T_{2}(\lambda_{\mathrm{s}},\lambda_{\mathrm{i}};\psi)\widetilde{G}^{(2)}(\lambda_{\mathrm{s}},\lambda_{\mathrm{i}})$; here, the biphoton transfer function is $T_{2}(\lambda_{\mathrm{s}},\lambda_{\mathrm{i}};\psi)=T(\lambda_{\mathrm{s}};\psi)T(\lambda_{\mathrm{i}};\psi)$, $T(\lambda_{\mathrm{s}};\psi)$ is the single-photon omni-resonant transfer function plotted in Fig.~\ref{Fig:EntPhotOmniResData}e (taken from Fig.~\ref{Fig:SinglePhotonData}c), and $\widetilde{G}^{(2)}(\lambda_{\mathrm{s}},\lambda_{\mathrm{i}})$ is the SPDC biphoton spectrum (taken from Fig.~\ref{Fig:EntangledPhotonSetup}c). We select three values for $\psi$, and we plot in Fig.~\ref{Fig:EntPhotOmniResData}f-h the biphoton transfer function $T_{2}(\lambda_{\mathrm{s}},\lambda_{\mathrm{i}};\psi)$ along with the transmitted biphoton spectrum $\widetilde{G}_{\mathrm{FP}}^{(2)}(\lambda_{\mathrm{s}},\lambda_{\mathrm{i}};\psi)$. For $\psi=0^{\circ}$ (Fig.~\ref{Fig:EntPhotOmniResData}f), the transfer function is narrow and does not overlap with the degenerate wavelength $\lambda_{\mathrm{o}}=810$~nm. Consequently, the transmitted biphoton spectrum vanishes everywhere $\widetilde{G}_{\mathrm{FP}}^{(2)}(\lambda_{\mathrm{s}},\lambda_{\mathrm{i}};0^{\circ})\approx0$. For $\psi=10^{\circ}$ (Fig.~\ref{Fig:EntPhotOmniResData}g), the transfer function is also narrow, but it now overlaps with the degenerate wavelength $\lambda_{\mathrm{o}}=810$~nm. The transmitted biphoton spectrum in this case allows for the biphoton spectrum $\widetilde{G}_{\mathrm{FP}}^{(2)}(\lambda_{\mathrm{s}},\lambda_{\mathrm{i}};10^{\circ})$ in the vicinity of the degenerate condition to be admitted to the cavity. In contrast, when $\psi=59^{\circ}$ (Fig.~\ref{Fig:EntPhotOmniResData}h), the transfer function is spectrally flat across the bandwidth $\Delta\lambda$ centered at $\lambda_{\mathrm{o}}=810$~nm. Consequently, the transmitted biphoton spectrum is identical to the SPDC biphoton spectrum $\widetilde{G}_{\mathrm{FP}}^{(2)}(\lambda_{\mathrm{s}},\lambda_{\mathrm{i}};59^{\circ})\approx\widetilde{G}^{(2)}(\lambda_{\mathrm{s}},\lambda_{\mathrm{i}})$. In this biphoton omni-resonant configuration, the entire bandwidth of both photons is admitted to the cavity, and the biphoton spectral structure is preserved.

\section*{Discussion}

A different perspective can elucidate the concept of omni-resonance. A conventional FP resonance stems from longitudinal phase-matching that ensures constructive interference of recycled light associated with a single transverse mode; namely the round trip phase must be an integer multiple of $2\pi$. The single-transverse-mode restriction results in only a single wavelength satisfying this condition and thus resonating for each longitudinal mode. However, this resonant wavelength changes for different transverse modes even when the longitudinal mode is held fixed. In the case of a planar FP cavity, the transverse modes are plane waves associated with different incident angles. Omni-resonance can therefore be viewed as a mapping between wavelengths and transverse cavity modes while maintaining a fixed longitudinal mode. The required precision of this mapping is determined by the cavity finesse \cite{Shiri22OL}: higher cavity finesse necessitates a precise association between wavelength and incident angle (i.e., low spectral uncertainty).

Omni-resonance in a planar FP cavity is achieved through pre-conditioning the optical field prior to incidence by introducing judicious AD that guarantees each wavelength in the spectrum to resonate in the cavity \cite{Shabahang17SR}. Because only the field is modulated while the cavity itself is not modified, the underlying cavity finesse is unchanged. The resulting achromatic resonance -- associated with a single underlying cavity longitudinal resonance -- can have in principle a bandwidth that is dramatically broader than the narrow resonant linewidth determined by the cavity finesse. For the FP resonance of order $m=29$ in the cavity used in our experiment, the resonant bandwidth of the associated achromatic resonance can extend in principle to $\Delta\lambda_{29}=278$~nm, compared to the resonant linewidth of $\delta\lambda\approx0.3$~nm. We have nevertheless exploited only a fraction $\Delta\lambda\approx20$~nm of this available bandwidth to minimize the complexity of the required pre-conditioning system. To achieve omni-resonance over a bandwidth of $\Delta\lambda\approx20$~nm at oblique incidence, only linear AD is required, which is readily provided by a diffraction grating. Increasing the bandwidth further or operating at near-normal incidence on the cavity requires a more complex AD synthesis setup. To realize omni-resonance close to normal -- rather than oblique -- incidence requires highly nonlinear AD of the form $\varphi_{m}(\lambda)\propto\sqrt{\lambda_{m}-\lambda}$, which cannot be provided by gratings or any other single-surface optic. In this case, a more complex optical system is utilized such as that in Ref.~\cite{Shiri20OL}, which is the system developed for synthesizing space-time wave packets \cite{Yessenov22AOP,Hall24OE}. Moreover, recent developments have extended AD synthesis to two transverse dimensions (conical AD \cite{Yessenov22NC,Yessenov22OL,Yessenov25JOSAA}), which can also be useful in realizing omni-resonance in a planar FP cavity. Further extending the bandwidth for omni-resonance at oblique incidence also requires nonlinear AD; see Ref.~\cite{Hall25LPR} where we demonstrate $>100$-nm omni-resonant bandwidth in the visible. Exploiting the full bandwidth $\Delta\lambda_{m}$ available to each longitudinal resonance in the omni-resonant configuration may allow for even attosecond laser pulses to couple to a cavity.

A key motivation for enabling the coupling of broadband quantum states of light (single-photon and entangled-photon states here) to an FP cavity is to achieve resonant field enhancement without spectral filtering and while preserving spectral correlations -- a crucial requirement for applications such as broadband coherent perfect absorption and few-photon nonlinear effects. Indeed, coupling of frequency-entangled photons to a cavity is the ultimate test for any finesse-preserving, linewidth-broadening mechanism for two reasons: the bandwidth of such photons is typically large and their frequencies are anti-correlated. Except in the degenerate condition (when the two photons have the same wavelength), when one photon couples to a longitudinal resonance, the other photon will be rejected by the cavity. Coupling the pair of frequency-anticorrelated entangled photons is thus unambiguous proof of the linewidth broadening in the omni-resonant configuration while preserving their continuous spectra. Note that the pre-conditioning optical system is linear, and both the signal and idler photon are subjected to the same AD modulation. Although the signal and idler photons have anti-correlated frequencies, each frequency in nevertheless directed to the FP cavity at the designated angle, do that both photons resonate simultaneously across their entire bandwidth (which is limited only by the design of the pre-conditioning system).

The results reported here suggest several avenues for future experiments. An interesting avenue is to implement omni-resonance for squeezed light, which may be useful in enhancing the detection of gravitational waves \cite{Aasi13NP,Ganapathy23PRX,Jia24Science}. Another potential line of investigation is broadband resonant quantum-enhanced spectroscopy \cite{Dorfman16PMP}, entangled two-photon absorption \cite{Varnavski20JACS}, and frequency conversion while maintaining frequency-entanglement \cite{Kumar90OL,Huang92PRL,McGuinness10PRL,Ates12PRL,Huang13OL,Lefrancois15PRA,Wang23PRA}. Coupling broadband single-photon omni-resonance with coherent perfect absorption can help yield ultrathin single-photon detectors with high efficiency. Moreover, omni-resonance can enable quantum information processing in the spectral domain by implementing programmable wavelength-selective omni-resonance in which the resonating spectrum can be sculpted at will \cite{Shiri20APLP}. Indeed, integrating beam-shaping strategies with FP cavities \cite{Slobodkin22Science} unlocks fresh opportunities for single-photon applications in quantum communication, sensing, and nonlinear optics. Our results therefore have broad implications for quantum optics and photonics, especially in photon-starved applications where omni-resonance can enable the use of high-finesse cavities with broadband quantum states of light. More broadly, the work reported here is an addition to the recent progress in utilizing spatiotemporally structured light in quantum optics.

\clearpage
\section*{Methods}

\noindent\textbf{FP cavity.} The planar FP cavity comprises two multilayer dielectric Bragg mirrors, each comprising 5 bilayers of SiO$_{2}$ and TiO$_{2}$ (of thicknesses 138~nm and 88~nm, and refractive indices 1.45 and 2.37, respectively, at a wavelength of 810~nm), each providing reflectivity of $R=0.97$. The two mirrors sandwich a thin silica layer of length $\approx9.88$~$\mu$m and refractive index $1.45$ at a wavelength of 810~nm. All the layers are deposited with e-beam evaporation. The resulting FP cavity has a finesse of $\mathcal{F}\approx 100$ and a quality-Factor $Q\approx3670$. The cavity resonant linewidth is $\approx0.3$~nm and the FSR is $\approx22$~nm. We used the 29$^{\mathrm{th}}$-order resonant mode of the cavity, which requires introducing angular dispersion with linear coefficient $\gamma_{29}\approx0.23^\circ$/nm to achieve omni-resonance when tilting the FP cavity an angle $\psi=59^{\circ}$ with the normal for $\lambda=810$~nm.

\noindent\textbf{Setup for introducing AD to achieve omni-resonance.} A collimated field is incident on a blazed grating with 1200 lines/mm at an incident angle of $\alpha\approx40.3^{\circ}$. The grating produces an AD coefficient of $\gamma\approx0.07^{\circ}$/nm for the first diffraction order produced at an angle $19.2^{\circ}$ centered at 810~nm. The diffracted light is then imaged through a de-magnification system consisting of two lenses L$_{1}$ and L$_{2}$ (of focal lengths $f=200$ and 55~mm, respectively). This increases the AD produced by the grating to reach the target value $\gamma=\gamma_{29}\approx0.23^{\circ}$/nm required for omni-resonance. The cavity is placed in the image plane.

\noindent\textbf{Entangled-photon source.} A monochromatic laser (Coherent, Cube) at a wavelength of 405~nm, power of 50~mW, and $2w$ beam width of 1 mm is focused into the crystal by a 100-mm lens as a pump for a 3-mm-thick beta-barium borate (BBO) nonlinear crystal (NLC) to maximize coupling into a SMF. The NLC cut angle is $\approx29^{\circ}$, which produces collinear degenerate type-I SPDC. The pump after the NLC is eliminated using a 40-nm-bandwidth bandpass filter centered at 800~nm (Thorlabs, FBH800-40) which restricts the correlation bandwidth from 800~nm to 820~nm. The system is aligned using an SLD of bandwidth 30~nm centered at 810~nm (Thorlabs, SLD810S), which is combined with the pump laser into a single path via a dichroic mirror. 

\begin{figure*}[t!]
\centering
\includegraphics[width=17.6cm]{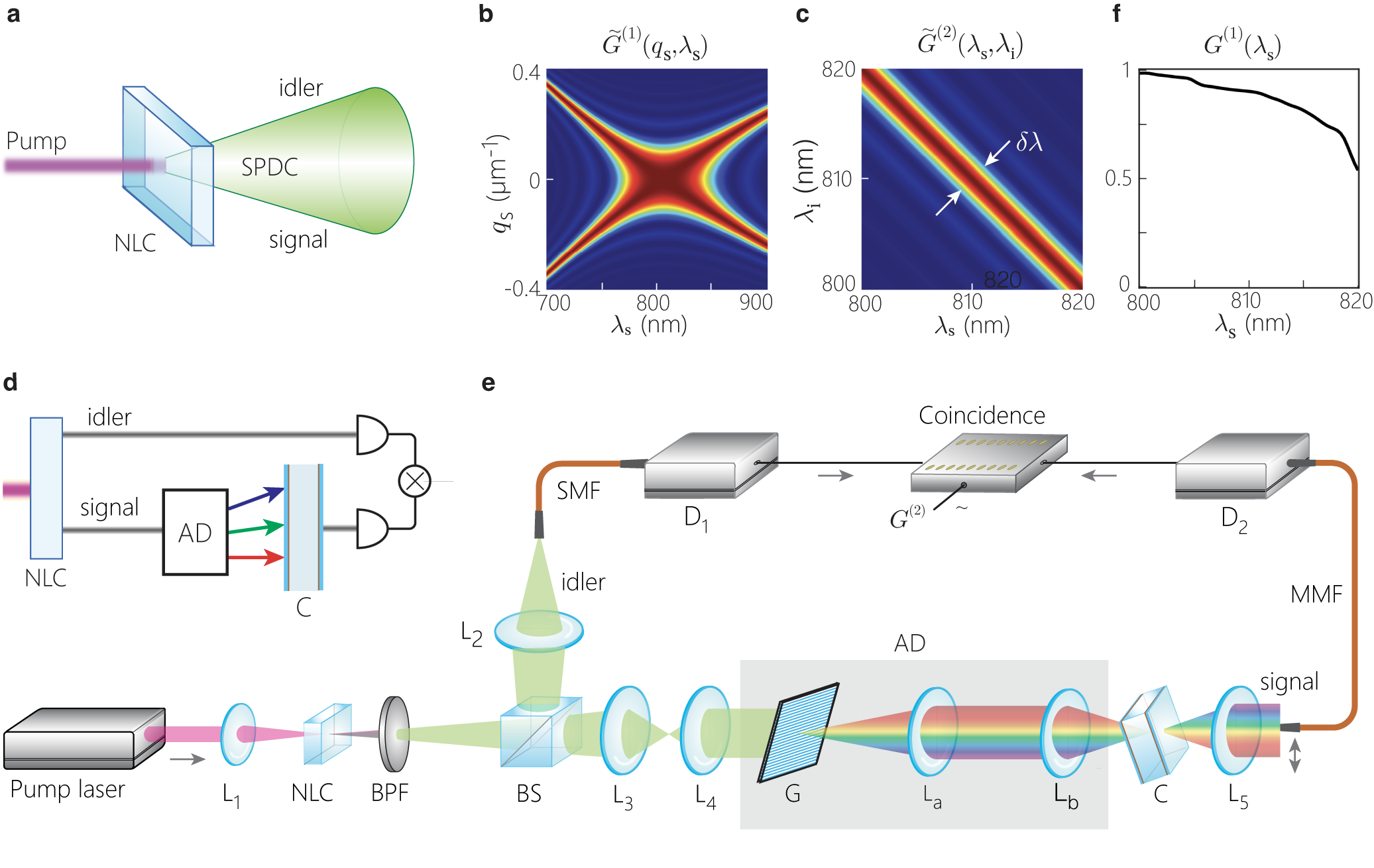}
\caption{\textbf{Configuration for single-photon omni-resonance.} (a) Schematic of the angular emission pattern from SPDC illustrating the signal and idler photons emitted on opposite sides of the emission cone; NLC: nonlinear crystal. (b) The spatiotemporal spectrum $\widetilde{G}^{(1)}(q_{\mathrm{s}},\lambda_{\mathrm{s}})$ for the signal photon. (c) The biphoton spectrum $\widetilde{G}^{(2)}(\lambda_{\mathrm{s}},\lambda_{\mathrm{i}})$ showing the frequency-anticorrelation structure, calculated for a 3-mm-long BBO crystal. A spectral uncertainty of $\delta\lambda\approx2.4$~nm is introduced into the frequency anticorrelation as expected from the finite spectral resolution of the measurement system; see Fig.~\ref{Fig:EntangledPhotonSetup}c. (d) Schematic of the heralded measurements. (e) Setup for omni-resonance measurement configuration. L$_{1}$-L$_{5}$, L$_{\mathrm{a}}$, L$_{\mathrm{b}}$: Lenses; BPF: bandpass filter; BS: beam splitter; SMF: single-mode fiber; MMF: multimode fiber; D$_{1}$, D$_{2}$: single-photon detectors. (f) The measured single-photon spectrum $\widetilde{G}^{(1)}(\lambda_{\mathrm{s}})$ after removing the cavity from the setup.}
\label{Fig:heralded_setup}
\end{figure*}

\begin{figure*}[t!]
\centering
\includegraphics[width=17.6cm]{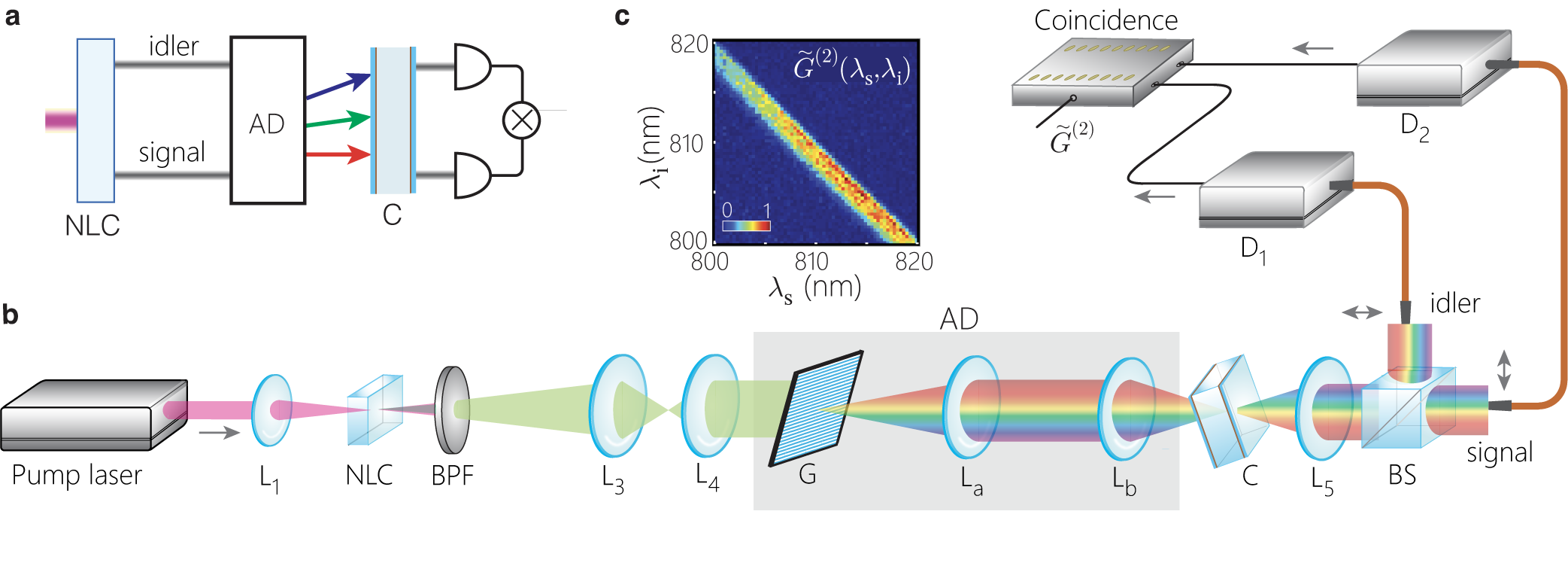}
\caption{\textbf{Configuration for entangled-photon omni-resonance.} (a) Conceptual schematic for biphoton omni-resonance. (b) Setup for observing biphoton omni-resonance. The optical components are the same as in Fig.~\ref{Fig:heralded_setup}e. The beam splitter that precedes the omni-resonant pre-conditioning system in Fig.~\ref{Fig:heralded_setup}e is removed, and a beam splitter is placed instead after the FP cavity to separate the signal and idler and measure their joint spectrum. Two multimode fibers are scanned, one corresponding to $\lambda_{\mathrm{s}}$ and the other to $\lambda_{\mathrm{i}}$, to capture the biphoton spectrum $\widetilde{G}_{\mathrm{FP}}^{(2)}(\lambda_{\mathrm{s}},\lambda_{\mathrm{i}})$. (c) The measured SPDC biphoton spectrum $\widetilde{G}^{(2)}(\lambda_{\mathrm{s}},\lambda_{\mathrm{i}})$ as produced by the NLC, which is recorded when the cavity is removed from the setup. The corresponding calculated biphoton spectrum is plotted in Fig.~\ref{Fig:heralded_setup}c.}
\label{Fig:EntangledPhotonSetup}
\end{figure*}

\noindent\textbf{Single-photon heralding.} In the system illustrated in Fig.~\ref{Fig:heralded_setup}e, a beam splitter after the NLC separates the signal and idler photons. The signal photon is directed to the omni-resonant system, while the idler is coupled into a single-mode fibre (Thorlabs, P1-780A-FC-1; mode field diameter $5 \pm 0.5~\mu$m, NA 0.13). An imaging system with $\sim10\times$ magnification -- comprising lenses L$_{3}$ and L$_{4}$ of focal lengths 50~mm and 500~mm, respectively -- relays the NLC output to the grating plane in the omni-resonant system. A 2$f$ system (lens L$_{2}$, $f=50$~mm) focuses the idler photon to a single-mode fiber. After transmission through the cavity, the signal photon is spatially dispersed by a lens ($f = 50$ mm, 2$f$ geometry), and the spectrum is mapped using a multimode fiber (Thorlabs, M43L02; 105~$\mu$m core, NA 0.22) mounted on a linear stage (Melles Griot, NanoMotion Motors). The single-mode (idler) and multimode (signal) fibers are coupled to single-photon detectors (PerkinElmer, SPCM-AQR-15-FC), and coincidence events are recorded using a time-tagging module (Swabian Instruments, Time Tagger Ultra; 42~ps resolution, 2~ns window).

\noindent\textbf{Two-photon, dual-frequency coincidence measurements.} In this configuration, the beam splitter before the cavity is removed. Instead, both photons—signal and idler—enter the omni-resonant system and traverse the cavity together. After the cavity, a lens ($f = 50$ mm) in a 2$f$ configuration spatially resolves their spectral content. A beam splitter placed after this lens separates the photons into two distinct paths. Each path is coupled into a multimode fiber (Thorlabs, M43L02; 105~$\mu$m core, NA = 0.22) mounted on independent linear motorized stages (Melles Griot, NanoMotion Motors) to scan across the resolved spectrum. Both multimode fibers are connected to single-photon detectors (PerkinElmer, SPCM-AQR-15-FC), and coincidence counts are recorded using the same time-tagging module described above.

\begin{figure*}[t!]
\centering
\includegraphics[width=17.6cm]{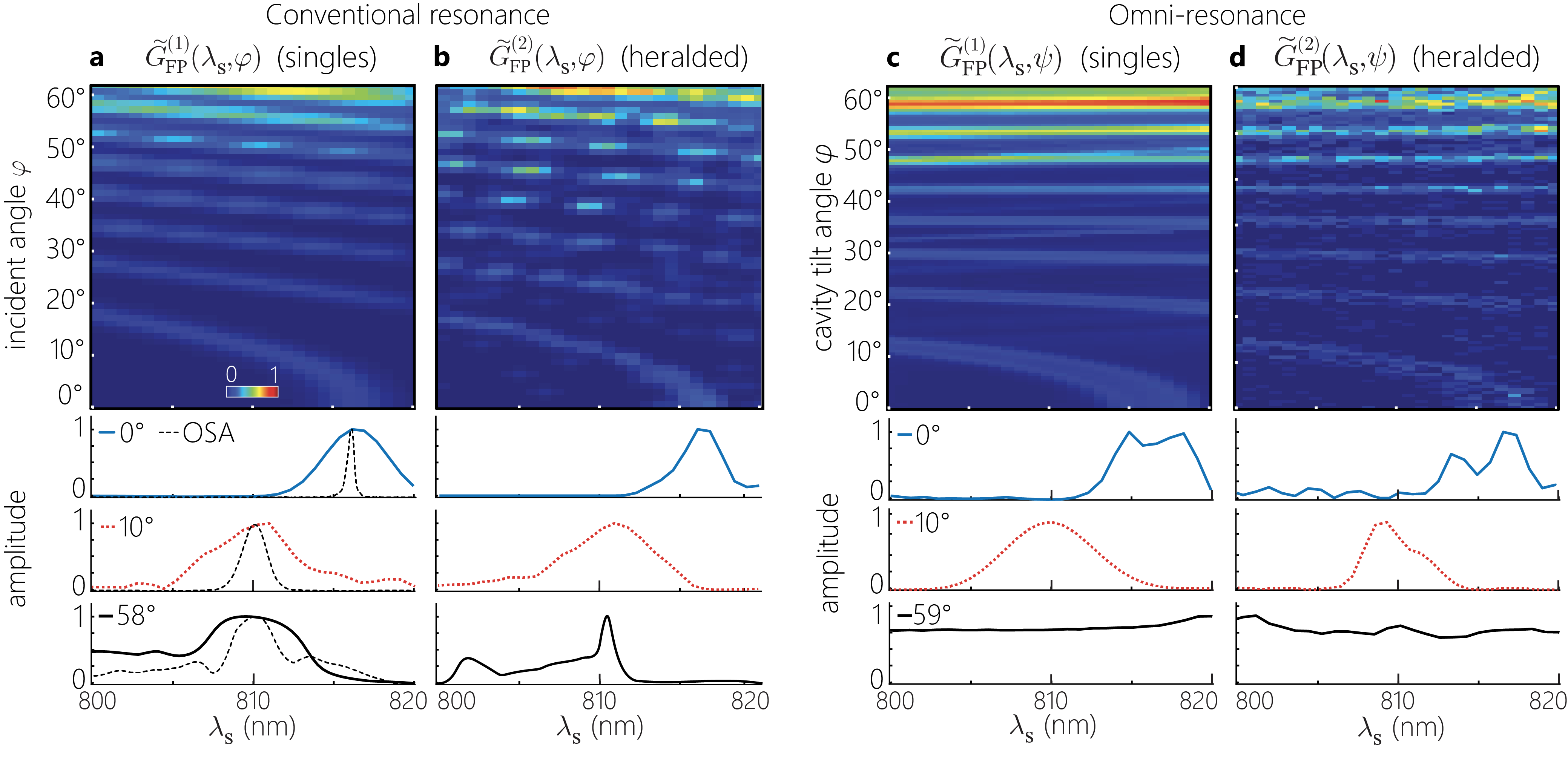}
\caption{\textbf{Demonstration of broadband single-photon omni-resonance in a planar FP cavity.} (a) Measured singles $\widetilde{G}_{\mathrm{FP}}^{(1)}(\lambda_{\mathrm{s}},\varphi)$ and (b) heralded $\widetilde{G}_{\mathrm{FP}}^{(2)}(\lambda_{\mathrm{s}},\varphi)$ rates for the signal photon after traversing the FP cavity in the conventional resonant configuration. The FP cavity is placed before the pre-conditioning system in Fig.~\ref{Fig:heralded_setup}e. A collimated single-photon state is incident on the FP cavity. The bottom panels are sections through the measurements at $\varphi=0^{\circ}$, $10^{\circ}$, and $58^{\circ}$. The dashed curves are measurements using a strong classical source acquired with an optical spectrum analyzer and not the measurement setup in Fig.~\ref{Fig:heralded_setup}e, and are provided as a reference for comparison. (c) Measured singles $\widetilde{G}_{\mathrm{FP}}^{(1)}(\lambda_{\mathrm{s}},\psi)$ and (d) heralded $\widetilde{G}_{\mathrm{FP}}^{(2)}(\lambda_{\mathrm{s}},\psi)$ rates for the signal photon in the omni-resonant configuration. The FP cavity is placed after the pre-conditioning system as shown in Fig.~\ref{Fig:heralded_setup}e, so that the single-photon state incidence on the FP cavity incorporates AD. The bottom panels are sections through the measurements at $\psi=0^{\circ}$, $10^{\circ}$, and $59^{\circ}$, the latter of which is the target omni-resonant condition.}
\label{Fig:SinglePhotonData}
\end{figure*}

\begin{figure*}[t!]
\centering
\includegraphics[width=17.6cm]{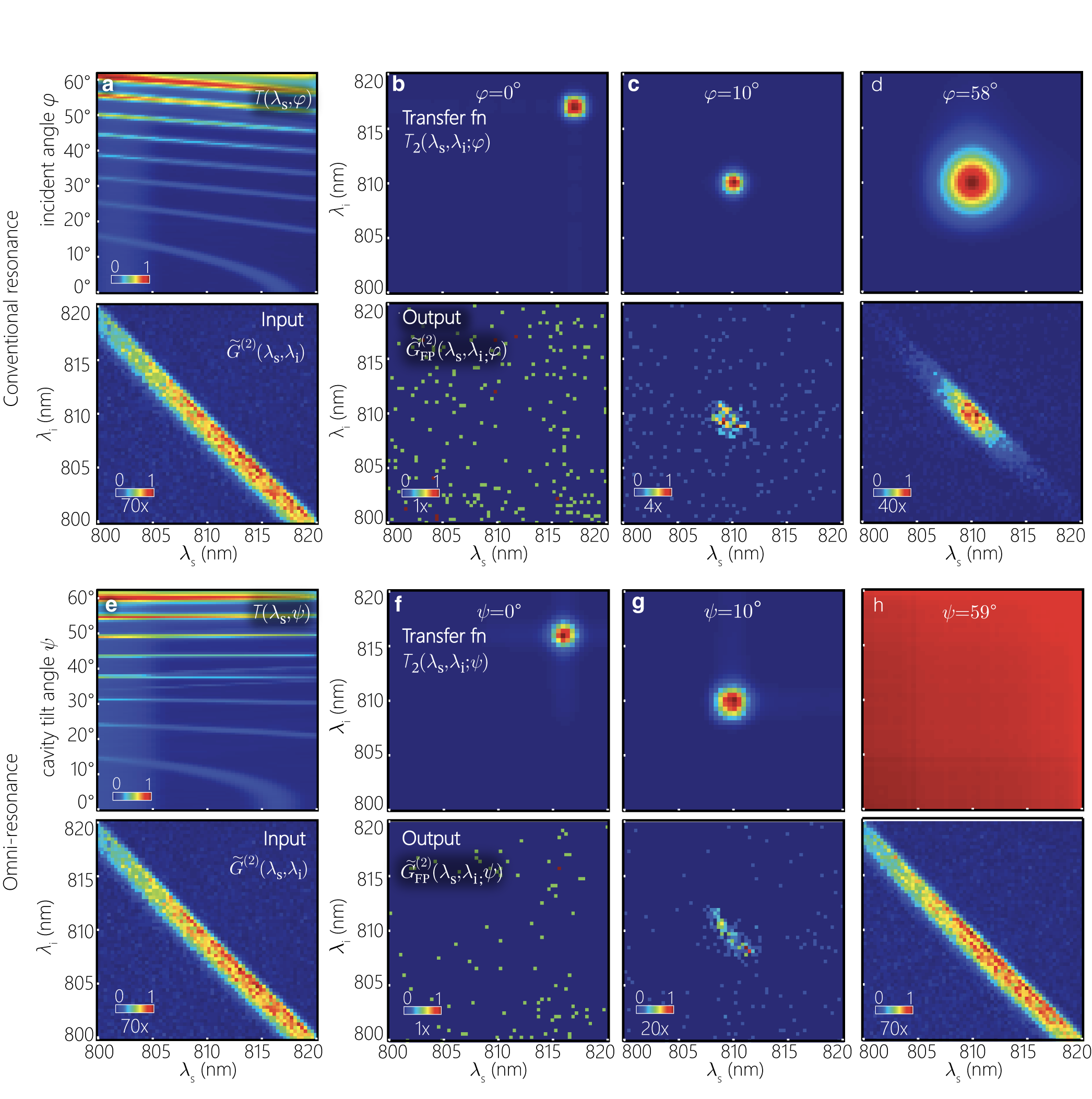}
\caption{\textbf{Demonstration of broadband entangled-photon omni-resonance in a planar FP cavity.} (a) Measured single-photon transfer function $T(\lambda,\varphi)$ in a conventional resonant configuration in the top panel (from Fig.~\ref{Fig:SinglePhotonData}a), and the measured biphoton spectrum $\widetilde{G}^{(2)}(\lambda_{\mathrm{s}},\lambda_{\mathrm{i}})$ in the bottom panel (from Fig.~\ref{Fig:EntangledPhotonSetup}c). (b-d) The top panel is the biphoton spectral transfer function $T_{2}(\lambda_{\mathrm{s}},\lambda_{\mathrm{i}};\varphi)=T(\lambda_{\mathrm{s}};\varphi)T(\lambda_{\mathrm{i}};\varphi)$ extracted from (a), and the bottom panel is the measured output biphoton spectrum $\widetilde{G}^{(2)}_{\mathrm{FP}}(\lambda_{\mathrm{s}},\lambda_{\mathrm{i}};\varphi)=T_{2}(\lambda_{\mathrm{s}},\lambda_{\mathrm{i}};\varphi)\widetilde{G}^{(2)}(\lambda_{\mathrm{s}},\lambda_{\mathrm{i}})$ for (b) $\varphi=0^{\circ}$, (c) $10^{\circ}$; and (d) $58^{\circ}$. (e) Measured single-photon omni-resonant transfer function $T(\lambda,\psi)$ in the top panel (from Fig.~\ref{Fig:SinglePhotonData}c) and the measured biphoton spectrum $\widetilde{G}^{(1)}(\lambda_{\mathrm{s}},\lambda_{\mathrm{i}})$ in the bottom panel (from Fig.~\ref{Fig:EntangledPhotonSetup}c). (f-h) The top panel is the biphoton spectral transfer function $T_{2}(\lambda_{\mathrm{s}},\lambda_{\mathrm{i}};\psi)=T(\lambda_{\mathrm{s}};\psi)T(\lambda_{\mathrm{i}};\psi)$ extracted from (e), and the bottom panel is the measured output biphoton spectrum $\widetilde{G}^{(2)}_{\mathrm{FP}}(\lambda_{\mathrm{s}},\lambda_{\mathrm{i}};\psi)=T_{2}(\lambda_{\mathrm{s}},\lambda_{\mathrm{i}};\psi)\widetilde{G}^{(2)}(\lambda_{\mathrm{s}},\lambda_{\mathrm{i}})$ for (b) $\psi=0^{\circ}$, (c) $10^{\circ}$; and (d) $59^{\circ}$.}
\label{Fig:EntPhotOmniResData}
\end{figure*}

\noindent\textbf{Classical measurements of the conventional resonance response.} A collimated beam of diameter 1~mm from a superluminescent diode (Thorlabs, SLD810S) of bandwidth 30~nm centered at 810~nm is directed to the cavity at three incident angles: $\varphi=0^{\circ}$, $10^{\circ}$, and $58^{\circ}$. The transmitted field is focused by a lens of focal length $f=50$~mm into a multimode fiber (Thorlabs, M43L02), which is coupled to an optical spectrum analyzer (Ando AQ6317B) to record the spectrum. The measurements are plotted in the botton panels in Fig.~\ref{Fig:SinglePhotonData}a as dashed curves.

\clearpage
\bibliography{diffraction}

\vspace{6mm}
\noindent\textbf{Acknowledgements}
\noindent
This work was funded by the US Office of Naval Research (ONR) through MURI award N00014-20-1-2789.

\vspace{3mm}
\noindent
\textbf{Author contributions}
\noindent
A.F.A. conceived the idea; B.L.T. and A.F.A. developed the theory and designed the experiment; B.L.T. conducted the experiment and the data analysis. All the authors contributed to the interpretation of the results and manuscript writing. The project was supervised by A.F.A. and B.E.A.S. 

\vspace{3mm}
\noindent\textbf{Competing interests}
\noindent
The authors declare no competing interests.

\vspace{3mm}
\noindent
\textbf{Correspondence} and requests for materials should be addressed to Ayman Abouraddy.


\clearpage

\clearpage

\clearpage

\clearpage

\end{document}